\documentclass[12pt,thmsa]{article}
\usepackage{amsfonts}

\input{tcilatex}
\begin{document}

\author{Karl-Georg Schlesinger \qquad \\
Erwin Schr\"{o}dinger Institute for Mathematical Physics\\
Boltzmanngasse 9\\
A-1090 Vienna, Austria\\
e-mail: kgschles@esi.ac.at}
\title{Some notes on string theory}
\date{}
\maketitle

\begin{abstract}
We start from a quantum computational principle suggested for string theory
in a previous paper and discuss how it might lead to a dynamical principle
implying the correct classical gravitational limit. Besides this, we briefly
look at some structural properties of moduli space.
\end{abstract}

\section{Introduction}

In \cite{Sch} we suggested a quantum computational principle for string
theory which was formulated as follows:

\bigskip

\textbf{Principle: }

\begin{enumerate}
\item  \textit{All} physical systems should be amenable to a \textit{real}
time simulation on a quantum computer and quantum computers should be
described as physical systems by deformation quantization of classical
Turing machines.

\item  The observable quantities of the world should be those which can be
determined by observation of quantum computers on quantum computers.
\end{enumerate}

\bigskip

\begin{remark}
We should say more precisely at this point what we mean by the assumption
that the quantum Turing machine should be describable by deformation
quantization. For our purpose, here, it is sufficient to assume that the
quantum Turing machine can be described by a quantum field theory, the
arithmetic content of the Feynman diagrams of which is equivalent to the
data of the weights appearing in deformation quantization. The work of
Freedman et al. (see \cite{FLW}) shows that, indeed, one can describe
quantum computation in terms of low dimensional quantum field theory (this
description being equivalent to the one in terms of spin systems plus an
automatic inclusion of error correction). The quantum field theory used in 
\cite{FLW} is determined by a quantum group and $q$-deformation is known to
appear as a special case of deformation quantization (see \cite{MSSW}).
\end{remark}

It was discussed there that the first part of this principle leads to the
algebra $P_{\Bbb{Z},Tate}$ as a fundamental object of the theory and
therefore - under the assumption of the partly conjectural scenario of \cite
{Kon 1999} - to the Grothendieck-Teichm\"{u}ller group $GT$ as a fundamental
symmetry. The second part of the principle implies a ladder of quantization
and, especially, a deformation quantization of $GT$ (or $P_{\Bbb{Z},Tate}$)
to be relevant for the - yet to be discovered - full fledged quantized
string theory. We will first discuss the semi-classical setting where the
(undeformed) $GT$ is relevant and come to the question of a deformation
quantization of the mathematical structures involved only later. In order
for the above principle to make sense, it is decisive that the
Grothendieck-Teichm\"{u}ller group as a symmetry plus possibly some natural
requirements should basically determine the structural elements and the
dynamical principle of the theory. We will therefore investigate in this
paper some general features a theory incorporating $GT$ as a symmetry should
have.

Throughout this paper, we will assume the scenario of \cite{Kon 1999} to be
valid.

\bigskip

\section{Dynamics}

The algebra $P_{\Bbb{Z},Tate}$ is the algebra of periods of mixed Tate
motives unramified over $\Bbb{Z}$. Let $\mathcal{M}$ denote the moduli space
of mixed Tate motives unramified over $\Bbb{Z}$. There is a quite concrete
definition of motives as so called framed motives, the category of framed
motives being equivalent to a more abstractly defined category of mixed
motives (see \cite{Kon 1999}). A framed motive of rank $n$ is a matrix 
\[
A\in GL\left( n,P\right) 
\]
(where $P$ is the algebra of periods) such that 
\[
\Delta \left( A_{ij}\right) =\sum_{k,l}A_{ik}\otimes A_{kl}^{-1}\otimes
A_{lj} 
\]
with $\Delta $ the triple coproduct induced from the torsor which
corresponds to the isomorphism of Betti and de Rham cohomology. So, we can
understand $\mathcal{M}$ as the moduli space of framed motives with 
\[
A_{ij}\in P_{\Bbb{Z},Tate} 
\]
In \cite{Kon 1999} the moduli space $\mathcal{D}M$ of deformation
quantizations of a finite dimensional manifold $M$ is suggested to be a
proalgebraic variety which is a principal homogenous space of $GT$. This
space should have several structural features in common with $\mathcal{M}$
(as a moduli space of motives, $\mathcal{M}$ should also be a proalgebraic
variety and there should also be a natural action of $GT$). Since $\mathcal{M%
}$ is universal while $\mathcal{D}M$ depends on the given manifold $M$, one
suspects each $\mathcal{D}M$ to be a kind of ``representation'' of $\mathcal{%
M}$. We will make the assumption in the sequel that in this sense structural
features (like e.g. a metric) on $\mathcal{M}$ induce corresponding
structures on the ``representation'' $\mathcal{D}M$. One would suspect that
even beyond this there should be a natural duality between the class of all
``representations'' $\mathcal{D}M$ and $\mathcal{M}$, as e.g. known from the
Doplicher-Roberts theorem for the case of compact groups and from many other
examples.

\bigskip

\textbf{Conjecture:}

There should be a natural duality between $\mathcal{M}$ and some suitable
notion of a moduli space of all deformation quantizations (a kind of union
of the $\mathcal{D}M$).

\bigskip

We have not just taken the disjoint union of the $\mathcal{D}M$ because
presumably compactification of the individual proalgebraic varieties $%
\mathcal{D}M$ plays a role, here. In an appendix we will give a slightly
more involved argument in favour of the above conjecture.

The work of \cite{Kon 1997} and \cite{CF} shows that $\mathcal{D}M$ can
always be understood as a moduli space of two dimensional conformal field
theories. Since the Deligne conjecture (a proof of which is announced in 
\cite{Kon 1999}) plays a decisive role in the construction of $\mathcal{D}M$%
, the fact that the tangential structure of the extended moduli space of
string theory introduced by Witten is given by a total Hochschild complex
(see \cite{Wit}, \cite{Kon 1994}) suggests that, conversely, moduli spaces
of two dimensional conformal field theories should be structurally similar
to the spaces $\mathcal{D}M$. Also, the so called Cohomology Comparison
Theorem (CCT) of \cite{GeSch} leads to an argument to the effect that moduli
spaces of two dimensional conformal field theories should be interpretable
as moduli spaces of deformation quantizations. Together with the above
conjecture, we take the view in this paper that $\mathcal{M}$ should be
naturally a dual description of the moduli space $\mathcal{S}$ of classical
backgrounds of string theory. So, we suppose that one has the freedom to
switch between the two descriptions, e.g. when considering a dynamical
principle.

\bigskip

Accepting the suggested quantum computational principle, $\mathcal{M}$ is
the straightforward candidate for the state space of the theory. So,
assuming the duality between $\mathcal{M}$ and $\mathcal{S}$ to hold, the
principle would at least imply the correct kinematical arena for the theory.
It is then a decisive question if we get a natural suggestion for the
dynamics.

\begin{remark}
The conformal field theories corresponding to deformation quantizations in 
\cite{CF} are in no way restricted to the critical dimension of string
theory. So, it seems that one needs some additional requirements, like
critical dimension and the type of supersymmetry, in order to restrict $%
\mathcal{S}$ to really give superstring theory. This is what we meant above
when we spoke of $GT$ symmetry ```plus possibly some natural requirements''.
\end{remark}

\begin{remark}
We will exclusively discuss dynamics in a Euclidean sense in this paper. We
will not deal with the question of how to introduce a physically sensible
time parameter, here.
\end{remark}

As a first step, we will see that the $GT$ symmetry also fixes a Riemannian
metric on $\mathcal{M}$. Remember that $\mathcal{M}$ is a projective limit
of algebraic varieties of framed motives of fixed rank. Denote by $\mathcal{M%
}_n$ the moduli space of framed mixed Tate motives unramified over $\Bbb{Z}$
of rank $n\in \Bbb{N}$. $\mathcal{M}_n$ is, of course, a subvariety of $%
\left( P_{\Bbb{Z},Tate}\right) ^{n^2}$. On $\left( P_{\Bbb{Z},Tate}\right)
^{n^2}$ we can introduce the usual Euclidean distance. Let $\gamma _n$ be
the induced Riemannian metric on $\mathcal{M}_n$(i.e. the metric defined as
infimum of the length of curves in $\mathcal{M}_n$). The $\gamma _n$ induce
a Riemannian metric $\gamma $ on $\mathcal{M}$.

\begin{lemma}
The metric $\gamma $ is invariant under the action of $GT$ on $\mathcal{M}$
and is up to a normalization factor uniquely $_{}$determined by $GT$
invariance.
\end{lemma}

\proof%
Considering the action of $GT$ on $\mathcal{M}$ in terms of actions on the
components $\mathcal{M}_n$, we get a representation in terms of tensors $%
T_{i_1j_1i_2j_2}$ with $n^4$ coefficients in $P_{\Bbb{Z},Tate}$ acting on
framed motives $A_{ij}$ of rank $n$ by 
\[
\left( TA\right) _{i_1j_1}=T_{i_1j_1i_2j_2}A_{i_2j_2} 
\]
(here, and in the sequel, we apply the Einstein summation convention).

Since the image of an element of $\mathcal{M}_n$ under $T$ has to be an
element of $\mathcal{M}_n$, we have 
\[
\Delta \left( TA\right) _{i_1j_1}=\left( TA\right) _{i_1k_1}\otimes \left(
TA\right) _{k_1l_1}^{-1}\otimes \left( TA\right) _{l_1j_1} 
\]
Since $\Delta $ is an algebra morphism on $P_{\Bbb{Z},Tate}$, we have 
\begin{eqnarray*}
\Delta \left( TA\right) _{i_1j_1} &=&\Delta \left(
T_{i_1j_1i_2j_2}A_{i_2j_2}\right) \\
&=&\Delta \left( T_{i_1j_1i_2j_2}\right) \Delta \left( A_{i_2j_2}\right)
\end{eqnarray*}
i.e. 
\[
\Delta \left( TA\right) _{i_1j_1}=\Delta \left( T_{i_1j_1i_2j_2}\right)
\left( A_{i_2k_2}\otimes A_{k_2l_2}^{-1}\otimes A_{l_2j_2}\right) 
\]
On the other hand, 
\begin{eqnarray*}
&&\left( TA\right) _{i_1k_1}\otimes \left( TA\right) _{k_1l_1}^{-1}\otimes
\left( TA\right) _{l_1j_1} \\
&=&T_{i_1k_1i_2k_2}A_{i_2k_2}\otimes
A_{k_2l_2}^{-1}T_{k_2l_2k_1l_1}^{-1}\otimes T_{l_1j_1l_2j_2}A_{l_2j_2} \\
&=&\left( T_{i_1k_1i_2k_2}\otimes \left( T^{-1}\right)
_{k_1l_1k_2l_2}^T\otimes T_{l_1j_1l_2j_2}\right) \left( A_{i_2k_2}\otimes
A_{k_2l_2}^{-1}\otimes A_{l_2j_2}\right)
\end{eqnarray*}
where $^T$ denotes the transposed matrix on interpreting $T$ as a matrix -
with a first and a second double index - acting on $\left( P_{\Bbb{Z}%
,Tate}\right) ^{n^2}$. So, 
\[
\Delta \left( T_{i_1j_1i_2j_2}\right) =T_{i_1k_1i_2k_2}\otimes \left(
T^{-1}\right) _{k_1l_1k_2l_2}^T\otimes T_{l_1j_1l_2j_2} 
\]
holds on the span of the tangent spaces of $\mathcal{M}_n$.

Now, by the definition of the torsor underlying $\Delta $ and since $GT$ is
a quotient of the motivic Galois group, $T$ commutes with $\Delta $, i.e. 
\[
\Delta \left( T_{i_1j_1i_2j_2}A_{i_2j_2}\right) =\left(
T_{i_1k_1i_2k_2}\otimes T_{k_1l_1k_2l_2}\otimes T_{l_1j_1l_2j_2}\right)
\Delta \left( A_{i_2j_2}\right) 
\]
and we get 
\[
\Delta \left( T_{i_1j_1i_2j_2}\right) =T_{i_1k_1i_2k_2}\otimes
T_{k_1l_1k_2l_2}\otimes T_{l_1j_1l_2j_2} 
\]
on the span of the tangent spaces of $\mathcal{M}_n$. But since in the
orthogonal complement of the span of the tangent spaces of $\mathcal{M}_n$
we are free to make a choice for $T$ (under the restriction that 
\[
T\in GL\left( n^2,P_{\Bbb{Z},Tate}\right) 
\]
holds), we can assume that both equations for $\Delta \left(
T_{i_1j_1i_2j_2}\right) $ hold without restriction. But this means $T$ is an
orthogonal $n^2\times n^2$ matrix. In consequence, the metric on $\mathcal{M}%
_n$ which is invariant under the transformations $T$ is up to a
normalization factor uniquely determined and is given by $\gamma _n$.

\endproof%

\bigskip

Assuming the duality of $\mathcal{M}$ and $\mathcal{S}$, there should be a
unique $GT$ invariant Riemannian metric on $\mathcal{S}$, then. As is well
known, there is a natural metric on the moduli space of two dimensional
conformal field theories if one remembers that infinitesimal deformations of
conformal field theories are parametrized by local operators. The metric is
then simply defined by the two point function. So, in view of the above
uniqueness result, it would suffice to show $GT$ invariance of the two point
function metric $d$, in order to identify $\gamma $ and $d$ as dual to each
other.

Remember that we assumed above that $\mathcal{S}$ carries a transitive
action of $GT$, too. But, again, using the fact that infinitesimal
deformations of conformal field theories are parametrized by local
operators, we conclude that the Lie algebra of $GT$ (see \cite{Dri}) should
naturally act as a symmetry on two dimensional conformal field theories. But
if this is true, i.e. if the action of $GT$ on $\mathcal{S}$ induces a
universal symmetry property of two dimensional conformal field theories, the
two point function metric - as based on observables - would have to be
invariant.

There is a hint that the conclusion we have just drawn might indeed be
correct. The structure of the Connes-Kreimer Hopf algebra of renormalization
is in accordance with the arithmetic properties of deformation quantization
and of the Drinfeld associator, i.e. with the data of the
Grothendieck-Teichm\"{u}ller group (see \cite{CK}, \cite{Kon 1999}). So, it
seems that the symmetry properties behind the renormalization scheme might
be intimately linked to $GT$ invariance of moduli space.

\bigskip

A given metric canonically determines a dynamical principle by geodesic
motion. So, the suggested quantum computational principle has a natural
dynamical law corresponding to it as geodesic motion with respect to $\gamma 
$ on $\mathcal{M}$, respectively, $d$ on $\mathcal{S}$. For a quantum theory
a Klein-Gordon type equation is the counterpart of geodesic motion, i.e. for
quantized string theory one would anticipate a Klein-Gordon equation on $%
\mathcal{M}$ (or $\mathcal{S}$). The reader acquainted with work in quantum
gravity will notice that this is similar to the situation, there: The
Wheeler-De Witt equation is also formally a Klein-Gordon equation. Indeed,
this is not by accident, as we can see from the following lemma which shows
that the dynamics suggested here by abstract arguments has the correct
classical gravity limit.

\begin{lemma}
Geodesic motion with respect to the metric $d$ on $\mathcal{S}$ induces
Einstein dynamics in the classical limit of string theory if one restricts
to backgrounds which allow for a natural sclicing by a global time parameter.
\end{lemma}

\proof%
In the classical limit, the two point function metric has to induce a metric 
$\widetilde{d}$ on the space of Riemannian ($n-1$)-metrics (where $n=10$ for
string theory) which are Ricci flat backgrounds with a slicing by a global
time parameter. But by the properties of the two point function, $\widetilde{%
d}$ has to be a four index tensor field satisfying the locality requirement
of \cite{DeW}. But then there is a one parameter family of metrics
satisfying these criteria, only, and there is a natural value for the
parameter (see \cite{DeW}). So, basically there is a unique candidate for
the classical limit of the two point function metric, only. But the
Klein-Gordon equation with respect to this metric $\widetilde{d}$ is just
the Wheeler-De Witt equation of Einstein gravity.

\endproof%

\bigskip

In conclusion, there are indications that the suggested quantum
computational principle does not only lead to a natural choice for the
kinematical structure and for a dynamical principle for the theory but also
that the resulting dynamical principle has the correct classical limit in
the form of Einstein dynamics.

\bigskip

\begin{remark}
Though the above proof uses the fact that in the classical limit backgrounds
given by two dimensional conformal field theories go to Ricci flat metrics,
i.e. metrics satisfying the Einstein vacuum equations, the result is not
just a reformulation of this fact. The Ricci flat limit of two dimensional
conformal field theories refers to the consideration of single fixed
backgrounds. Our result shows that there is a dynamics on moduli space (i.e.
on the space of backgrounds) which is also compatible with Einstein
dynamics. A dynamics on moduli space is what one ultimately expects for the
yet to be discovered full formulation of string theory which gives the
motivation for studying qualitative properties and possible choices for
dynamical principles on moduli space. Especially, the above result shows
that the quantization of the suggested dynamics on moduli space would have
(in a formal sense) the Wheeler-De Witt equation as a special limit. While
the Ricci flat limit of conformal field theories shows that string theory
unifies Einstein gravity and quantum mechanics consistently on a
perturbative level, we get an indication, here, that string theory might
also have a limit in which it reproduces the nonperturbative approach to
canonical quantum gravity (see e.g. \cite{AL 1994}, \cite{AL 1996} and the
literature cited therein). A further elaboration on the role the
Grothendieck-Teichm\"{u}ller group plays in the structure of string theory
moduli space seems to be a promising candidate for the investigation of such
a limit (we are planing to undertake further work in this direction).
\end{remark}

\bigskip

\section{The implications of the second part}

So far, we have dealt only with the implications of the first part of the
principle. As we mentioned already, the second part implies that actually a
quantum deformation of $GT$ should be the relevant symmetry object.
Conformal field theories play the role of classical backgrounds around which
one can perturbatively expand in string theory (comparable to Ricci flat
metrics in Einstein gravity). One can give arguments that the inclusion of
non classical backgrounds (similar to general Riemannian 3-metrics for the
classical gravity case) should lead to quantum deformations of the
mathematical structures appearing (see \cite{GS}). Especially, deformations
of conformal field theories to models with noncommutative world sheets
should appear, here, as is indirectly also suggested by the work of \cite
{Gre}. A detailed discussion of a deformation of quantum group symmetries
(since these are linked to conformal field theories, i.e. to the undeformed
case in the sense of our present discussion) has been given elsewhere (see%
\cite{GS 2000}).

\bigskip

\section{$\zeta $-functions}

Consider the action functional $S$ generating the Klein-Gordon equation on $%
\mathcal{M}$ as the equation of motion. For physical reasons, we should
consider only action functionals which are at most second order in the
fields and their derivatives and do not contain higher derivatives of the
fields. But then the uniqueness result for the metric $\gamma $ implies that 
$S$ is - up to a normalization factor - the unique $GT$ invariant action
satisfying these criteria (a term linear in the fields would destroy $GT$
invariance, i.e. both terms - the one in the fields and the one in their
derivatives - have to be quadratic, similarly, there can be no mixed term in
fields and their derivatives). It is one of the beliefs of experts in the
theory of motives that $\zeta $-functions should be interpretable as a kind
of regularized volumes on configuration spaces of quantum fields (see \cite
{JKS}). The partition function is, of course, a natural candidate, then (an
additional argument in favour of an interpretation of $\zeta $-functions as
partition functions of quantum systems is provided by the work of \cite{Con}%
). Motivated by the uniqueness property of $S$, we suspect that the
partition function defined from $S$ is interpretable as $\zeta $-function of
the proalgebraic variety $\mathcal{M}$.

\bigskip

\begin{remark}
We should say that all differential geometric notions used in this section
are to be understood in a purely formal sense. To make the ideas sketched
precise, one would presumably have to use an approach replacing these by
techniques from algebraic geometry.
\end{remark}

\bigskip

\section{Conclusion}

We have investigated the implications of a quantum computational principle
for the kinematical and dynamical features of string theory. We have found,
especially, that such a principle might be able to determine to a
considerable extent the correct moduli space and is consistent with the
requirement that Einstein dynamics should appear as a suitable classical
limit.

\bigskip

\appendix 

\section{The question of duality of $\mathcal{M}$ and $\mathcal{S}$}

Suppose the connection between $GT$ and the Connes-Kreimer Hopf algebra can
be rigorously established. Then duality between $\mathcal{M}$ and $\mathcal{S%
}$ would come down to the following: First, every two dimensional quantum
field theory would have to be a representation of the renormalization
scheme. This is certainly true. Second, in the other direction we would need
a result showing that we can reconstruct the renormalization scheme from the
knowledge of all two dimensional quantum field theories. The historical
development gives a very strong argument in favour of such a reconstruction
possibility. After all, the renormalization scheme was not derived from
abstract logical arguments but was destilled from quantum field theories,
i.e. the historical development is itself a ``reconstruction'' (one should
better say ``construction'') process. This reconstruction possibility should
persist even if we restrict to two dimensional theories since one does not
have the feeling that we have to be free to consider arbitrary dimensions in
order to be able to discover the renormalization scheme. So, we suspect that
any fixed dimension should do.

Though the above argument relies on strong historical evidence, it is not a
proof, of course (historical evidence may deceive us in spite of being
strong). Can we do better? Since we are interested in a low dimensional
situation, only, we can try to use the framework of axiomatic quantum field
theory. Based on some of the concepts of this framework, we will try to give
a sketch for a more adequate mathematical argument, now. The framework of
algebraic quantum field theory nowdays makes use of the algebraic language
of categories. It is a well known experience in category theory that a
category of objects of some type often has a similar algebraic structure
than the objects themselves. It is precisely this feature of category theory
which we will see at work, here.

\bigskip

Let $\mathcal{Q}$ be the 2-category of $C^{*}$-quantum categories as
introduced in \cite{FK} where a quantum category is an abelian, semisimple,
finite, rigid, braided, monoidal category and a $C^{*}$-quantum category is
a quantum category with a compatible $*$ structure on it.

By the approach followed in \cite{Bae}, one can show that there is a tensor
product on $\mathcal{Q}$, turning it into a symmetric (weak) monoidal
2-category. In the same way, one proves the existence of a direct sum (note
that the finiteness condition of \cite{FK} implies what is called finite
dimensionality in \cite{Bae}). Also, by the results of \cite{Bae} one gets a
2-categorical version of rigidity for $\mathcal{Q}$.

The finiteness condition has some additional consequences since it implies
that up to isomorphism the morphisms in a $C^{*}$-quantum category can be
seen as finite tuples of matrices. But as a consequence of this, the
1-morphism classes of $\mathcal{Q}$ carry the structure of a $\Bbb{C}$%
-linear category and this linear structure is compatible with composition of
1-morphisms. Hence, $\mathcal{Q}$ carries the 2-categorical analog of the
structure of a $\Bbb{C}$-linear category. Besides this, by redoing the proof
that the category of finite dimensional vector spaces is abelian in a
categorified setting, one should find a 2-categorical counterpart of the
notion of an abelian category and get the result that $\mathcal{Q}$ has this
property.

In conclusion, we should find that $\mathcal{Q}$ is the 2-categorical
version of a $\Bbb{C}$-linear, abelian, rigid, symmetric monoidal category.
In the same way, one expects to find the 2-categorical analog of a $\Bbb{C}$%
-linear, exact, faithful, monoidal functor from $\mathcal{Q}$ to the
category of finite dimensional vector spaces. Now, assuming that there is a
2-categorical analog of the general reconstruction theorems for Hopf
algebras (see e.g. \cite{CP}), it would follow that there is a Hopf category 
$\mathcal{H}$ (in the sense of \cite{CrFr}) generating $\mathcal{Q}$ as its
category of either representations or corepresentations on finite
dimensional 2-vector spaces. By the symmetry of the monoidal structure, $%
\mathcal{H}$ would then have to be commutative or cocommutative. But this
means - now using the usual reconstruction theorems - $\mathcal{H}$ would
basically be a category of representations of some Hopf algebra \textbf{H}$.$
So, there should be a universal Hopf algebra $H$ having a representation on
every quantum field theory. Physically, the symmetries behind the
renormalization scheme are the only natural candidate for such a universal
Hopf algebra. We summarize this in the following conjecture:

\bigskip

\textbf{Conjecture:}

The Hopf algebra $H$ defined by $\mathcal{Q}$ should be equivalent to the
Connes-Kreimer Hopf algebra.

\bigskip

We have followed a purely algebraic approach to the question of duality
between the renormalization scheme and the general structure of quantum
field theory, here. But it should be possible to derive from such an
algebraic approach the more geometric duality between $\mathcal{M}$ and $%
\mathcal{S}$, too. In addition, this approach relies on a connection between
the Connes-Kreimer Hopf algebra and $GT$. But as we mentioned already, there
is growing evidence for such a connection to exist (see \cite{CK}, \cite{Kon
1999}).

\bigskip

\begin{remark}
The arguments given, here, are aiming at a direct proof of the assumed
duality in the sense of an algebraic reconstruction procedure.
Alternatively, one can from existing literature collect together a chain of
arguments which might also develop into a proof for the duality of $\mathcal{%
M}$ and $\mathcal{S}$. Concluding this appendix, we give a sketch of these
arguments, now: Surely, $\mathcal{M}$ and the Grothendieck-Teichm\"{u}ller
group, itself, are dual to each other. $\mathcal{M}$ is defined by
representations of $GT$ and, conversely, from $\mathcal{M}$ we get the
algebra $P_{\Bbb{Z},Tate}$ as the algebra of periods appearing in $\mathcal{M%
}$, and $P_{\Bbb{Z},Tate}$ determines $GT$ (where, as remarked at the
beginning, we assume the scenario of \cite{Kon 1999} to be valid). Starting
from the space $\mathcal{S}$, we remember that the fusion structure of two
dimensional conformal field theories is determined by quasi-triangular
quasi-Hopf algebras. So, $\mathcal{S}$ determines a certain moduli space of
quasi-triangular quasi-Hopf algebras. But then we can use Drinfeld's
original approach (see \cite{Dri}) to introduce $GT$ from the setting of
quasi-Hopf algebras. For the converse direction, we point, again, to the
increasing evidence for a deep relationship between $GT$ and the
Connes-Kreimer algebra. But, obviously, two dimensional conformal field
theories carry representations of the renormalization scheme as abstractly
displayed by the Connes-Kreimer algebra. So, we might get back $\mathcal{S}$
as a class of representations.
\end{remark}

\bigskip

\textbf{Acknowledgements:}

I would like to thank H. Grosse for numerous discussions and the Erwin
Schr\"{o}dinger Institute for Mathematical Physics, Vienna, for hospitality.

\end{document}